\documentclass[showpacs,preprintnumbers,amsmath,amssymb,twocolumn,nobibnotes,longbibliography,nofootinbib]{revtex4-2}
\usepackage[english]{babel}
\usepackage{mathtools}
\usepackage{graphicx}
\usepackage{mathrsfs}
\usepackage{amssymb}
\usepackage{xcolor}
\usepackage[colorlinks=true,citecolor=blue,urlcolor=blue]{hyperref}
\usepackage{amsmath}
\usepackage{bm}
\usepackage{physics}
\newcommand{\mysection}[1]{\textcolor{blue}{\textit{#1}.}}

\begin{document}

\title{Interplay of Rashba and valley-Zeeman splittings 
in weak localization of spin-orbit coupled graphene}

\author{L. E. Golub}
\affiliation{Physics Department, University of Regensburg, 93040 Regensburg, Germany}

\begin{abstract}
Weak localization theory is developed for graphene heterostructures with transition metal dichalcogenides and topological insulators where the Rashba and valey-Zeeman spin-splittings of the energy spectrum are large enough. The anomalous magnetoresistance in low fields caused by weak localization is calculated. It is shown that the valley-Zeeman splitting has no effect on weak localization in the absence of Rashba splitting but it results in the change of the magnetoconductivity sign in the Rashba-coupled graphene. Inter-valley scattering also affects the 
quantum correction to the conductivity
resulting in its sign reversal.
Analytical expressions are obtained for the anomalous magnetoconductivity at arbitrary relations between the Rashba and valley-Zeeman splittings as well as the inter-valley scattering rates.
\end{abstract}

\maketitle

\mysection{Introduction} 
Graphene 
proximitized by
strongly spin-orbit coupled materials attract a great deal of attention
due to its ability for spin
engineering~\cite{Zollner2025}. The most promising examples are  graphene heterostructures with topological insulators and transition metal dichalcogenides. In these systems, the Rashba splitting of Dirac fermions has an order of a few meV~\cite{Sierra2021}.
Another spin splitting in the absence of magnetic field, known as the valley-Zeeman splitting, is  present in graphene being opposite in the two valleys.
%
The valley-Zeeman splitting 
is also enhanced in the heterostructures~\cite{Naimer2023}. As a result, the Rashba and valley-Zeeman splittings are comparable and large enough to strongly affect quantum transport properties of the graphene heterostructures~\cite{Sun2023}. 

In low magnetic fields, quantum corrections to the conductivity are caused by weak localization (WL). The  anomalous magnetoresistance in graphene is positive in contrast to ordinary systems
because of the Berry phase of Dirac fermions in each valley equal to $\pi$. Therefore the interference has an opposite character and is known as weak antilocalization (WAL). However, the anomalous magnetoresistance depends crucially on the Rashba spin splitting~\cite{McCann2012,Golub2024}. In particular, a strong Rashba coupling results in the reversal of the sign of the conductivity correction, i.e. to WL.
At the same time, it is known that effective inter-valley scattering results in the transition from WAL to WL in graphene~\cite{McCann2006,Nestoklon2014}. 
%
%
As a result, both Rashba splitting and the inter-valley scattering change sign of the correction due to effect on the interference in the spin and valley spaces. Therefore, if they both are efficient, WAL takes place in the graphene heterostructures with negative magnetoconductivity in low fields~\cite{Wang2015,Wang2016,Zihlmann2018}. 

In this work we study real graphene heterostructures where all three ingredients -- Rashba spin splitting, valley-Zeeman splitting, and inter-valley scattering -- are present. We derive analytical expressions for the quantum corrections to the magnetoconductivity 
at arbitrary relations between them.

The Hamiltonian of the spin-orbit coupled graphene has the following form
\begin{equation}
\label{H}
\mathcal H = v (\xi \sigma_x p_x + \sigma_y p_y)+\lambda_{\rm R} (\xi\sigma_x s_y - \sigma_y s_x ) + \xi \lambda_{\rm VZ} s_z.
\end{equation}
Here $\bm p$ is momentum, $v$ is the Dirac fermion velocity, 
$x,y$ are coordinates in the graphene plane, 
$\xi = \pm$ enumerates the valleys,
$\lambda_{\rm R}$ and $\lambda_{\rm VZ}$ are the Rashba  spin-orbit and valley-Zeeman splittings, respectively.

The quantum correction to the conductivity is expressed via the Cooperon -- the amplitude of interference of two particles passing along the time-inversion coupled loops.
In the absence of valley-Zeeman splitting and spin- and valley-dependent disorders,  the Cooperon 
equals to $\mathcal L_0^{-1}$ where the operator $\mathcal L_0$ is given by~\cite{Golub2024}
\begin{equation}
\mathcal L_0 = {\hbar \over 4\abs{eB}}\qty[\qty(\bm q-{2\lambda_{\rm R}\over \hbar v}[\hat{\bm z}\times \bm S ])^2+{\Gamma_\phi\over D}].
\end{equation}
Here $\bm q$ is a generalized momentum of the pair of interferring particles
in the magnetic field $\bm B \parallel z$, $D$ is the diffusion coefficient, $\Gamma_\phi$ is the spin- and valley-independent dephasing rate, and 
$\bm S$ is the operator of the sum of angular momenta of two interfering states.
It is important that the operator $\mathcal L_0$ contains not only quadratic but also linear in $\bm q$ terms. They, also linear in $S_{x,y}$, couple the Cooperons in the spin triplet channel. This results in the expression for the magnetoconductivity in Rashba-coupled systems different from the classical Hikami-Larkin-Nagaoka (HLN) formula~\cite{ILP,Knap1996,Punnoose2006,Golub2024}.
 
The valley-Zeeman splitting gives an additional phase affecting the interference. In what follows we assume this splitting to be not too large, so that the parameter $\Delta = 2\lambda_{\rm VZ} \tau_{\rm tr}/\hbar$, where $\tau_{\rm tr}$ is the transport relaxation time, is much less than unity. 
However, the ratio
\begin{equation}
\Delta_\phi ={2\abs{\lambda_{\rm VZ}} \over \hbar \Gamma_\phi}
\end{equation}
might be arbitrary.
At $\Delta \ll 1$, the effect of the valley-Zeeman splitting 
is described by a term $-i\xi L_z \Delta$ added to 
$\mathcal L_0$. Here $\bm L$ is the operator of spin difference 
of two interfering states. It is different from the operator $\bm S$ 
because
the valley-Zeeman splitting is 
independent of momentum and, hence, does not change sign at the substitution $\bm p \to -\bm p$.
The operator $\bm L$ appears also in the WL problems for two-dimensional electrons in the in-plane magnetic field~\cite{Malshukov1997,Minkov2004,WAL_SST_Review} and exciton polaritons with an even in momentum longitudinal-transverse splitting~\cite{Glazov2008}.
A presence of both $\bm S$ and $\bm L$ operators results in a mixing of singlet and triplet spin channels of interference, which complicates the expressions for the magnetoconductivity.

We begin with a
calculation of the anomalous magnetoconductivity in the absence of inter-valley scattering. 
Then
we generalize the theory taking into account inter-valley scattering processes.

\mysection{Anomalous magnetoconductivity in the absence of inter-valley scattering}
If we ignore the inter-valley scattering, 
then the WL induced conductivity correction equals to a sum of two identical terms from one valley.
In each valley, WL is caused by an interference of Dirac fermions in different two-particle spin states. They are characterized be the total angular momentum $S$ and its projection onto $z$ axis $S_z$ and denoted as $t_1,t_0,t_{-1},s$, where
$t_m$ are the triplet channels with $S=1$, $S_z=m$, and $s$ is the singlet channel.
The operator $\mathcal L=\mathcal L_0-iL_z \Delta$  in the spin basis $t_1,t_0,t_{-1},s$ is 
the 4-rank matrix 
given by
\begin{equation}
\label{L_matrix}
 \mathcal L= \begin{pmatrix}
 & & & 0
\\  & \mathcal L_t & & ib_{\rm VZ}
\\  & & & 0
\\  0& ib_{\rm VZ}& 0& \epsilon
\end{pmatrix}, \quad b_{\rm VZ}
=\Delta_\phi b_\phi 
={\abs{\lambda_{\rm VZ}} \over 2\abs{e B}D}.
\end{equation}
Here 
$\epsilon=
\qty(q^2+\Gamma_\phi/ D)\hbar /(4\abs{eB})
$, and
the matrix $\mathcal L_t$ reads~\cite{Golub2024} 
\begin{equation}
\label{L_t_matrix_epsilon}
\mathcal L_t=
\begin{pmatrix}
\epsilon -1+b_{\rm R} & i\sqrt{2 b_{\rm R} n} &0 
\\ -i\sqrt{2 b_{\rm R} n} & \epsilon+2b_{\rm R} &i\sqrt{2 b_{\rm R} (n+1)} 
\\ 0 & -i\sqrt{2 b_{\rm R} (n+1)} & \epsilon+1+b_{\rm R}
\end{pmatrix},
\end{equation}
 \begin{equation}
 \label{b_Gamma}
b_i={\mathcal B_{i}\over \abs{B}}, \qquad \mathcal B_{i}= {\hbar \Gamma_{i} \over 4\abs{e}D}, 
\end{equation}
where 
$\Gamma_{\rm R}=2(\lambda_{\rm R}/\hbar)^2\tau_{\rm tr}$ is the Rashba-term induced Dyakonov-Perel spin relaxation rate, 
$\Gamma_{\rm VZ}=2\abs{\lambda_{\rm VZ}}/\hbar$,
and $n=\epsilon -b_\phi-1/2$.

Inverting the matrix~\eqref{L_matrix}
and calculating the conductivity correction, we obtain the WL induced magnetoconductivity $\Delta\sigma=\sigma(B)-\sigma(0)$ in the form~\cite{SM}
\begin{multline}
\label{sigma_no_iv}
\Delta\sigma=2\Delta\sigma_{\rm intra}, \quad
{\Delta\sigma_{\rm intra}(b_\phi) \over \sigma_0} 
=  - {1\over (b_\phi+b_{\rm R} )^2-1/4}
\\ - \sum_{m=1}^4 \qty[u_m \psi\qty(1/2 + b_\phi  - v_m)  - u_m^{(0)}\ln{\qty(b_\phi  - v_m^{(0)})}] .
\end{multline}
Here $\sigma_0=e^2/(2\pi h)$,  
 the common negative sign is caused by the Berry phase $\pi$ of Dirac fermions, 
and $\psi(y)$ is the digamma function.
The coefficients $v_{1\ldots 4}$ are the four roots of 
$\mathcal D(\epsilon)$ 
and $u_m = {\mathcal N(v_m)/\prod_{m'\neq m}(v_m-v_{m'})}$.
%
Here $\mathcal N(\epsilon)$ and $\mathcal D(\epsilon)$ are
the polynomials of the 3rd and 4th powers, respectively.
They are obtained from the equality
\begin{equation}
{\mathcal N(\epsilon)\over \mathcal D(\epsilon)} ={\rm Tr}
\qty(\mathcal E_4 \mathcal L^{-1}),
\end{equation}
where $\mathcal E_4={\rm diag}(1,1,1,-1)$, and the 4-rank matrix $\mathcal L$ 
is given by
Eq.~\eqref{L_matrix}.
The explicit expressions for $\mathcal N(\epsilon)$ and $\mathcal D(\epsilon)$ are given in Supplemental Material~\cite{SM}.
The coefficients $v_m^{(0)}$ and $u_m^{(0)}$ are the zero-field asymptotes of $v_m$ and $u_m$ calculated by 
passing to the limit $\epsilon \gg 1$ in the matrix $\mathcal L_t(\epsilon)$.
 
 The magnetoconductivity in the absence of the valley-Zeeman splitting has been calculated in Ref.~\cite{Golub2024}. In this case one has to invert the 3-rank matrix $\mathcal L_t$, see  Eqs.~\eqref{L_matrix} and~\eqref{L_t_matrix_epsilon}. The result is given by $\Delta\sigma=2\Delta\sigma_{\rm intra}$ with
 \begin{equation}
 \label{no_VZ}
{\Delta\sigma_{\rm intra} \over \sigma_0}\biggr|_{\lambda_{\rm VZ}=0} 
=
F(b_\phi)- \mathcal F_t(b_\phi,b_{\rm R}).
\end{equation}
Here 
$F(\beta)=\psi(1/2+\beta)-\ln{\beta}$
is the HLN function, and the spin triplet contribution is given by the function $\mathcal F_t$ derived in Ref.~\cite{Golub2024}, it is also presented in Supplemental Material~\cite{SM}.

If the Rashba splitting is absent, then the valley-Zeeman splitting has no effect on the anomalous magnetoconductivity. 
Indeed, in this case we have two independent spin subsystems in each valley with slightly different Fermi energies equal to $\epsilon_{\rm F}\pm \lambda_{\rm VZ} \approx \epsilon_{\rm F}$. This difference does not affect the magnetoconductivity which is independent of the Fermi energy. 
A lack of influence of $\lambda_{\rm VZ}$ on the conductivity correction in the absence of Rashba splitting
is also clear from Eqs.~\eqref{L_matrix} and~\eqref{L_t_matrix_epsilon}: the matrix $\mathcal L_t$ is diagonal at $b_{\rm R}=0$, and the $t_0$ and $s$ channels give equal contributions to conductivity of opposite signs at any value of $b_{\rm VZ}$ and cancel each other. The two rest spin channels, $t_{\pm 1}$, are not affected by the valley-Zeeman spitting making the conductivity correction independent of $\lambda_{\rm VZ}$. 
The magnetoconductivity in this limit is given by the HLN formula
 \begin{equation}
 \label{no_Rashba}
\Delta\sigma \bigr|_{\lambda_{\rm R}=0} 
=
-4\sigma_0F(b_\phi),
\end{equation}
where the factor `4' is due to two valleys and two spin interference channels.
This result can also be seen from Eq.~\eqref{no_VZ} by taking into account that $\mathcal F_t(b_\phi,0)=3F(b_\phi)$.

\begin{figure}[t]
	\centering \includegraphics[width=0.95\linewidth]{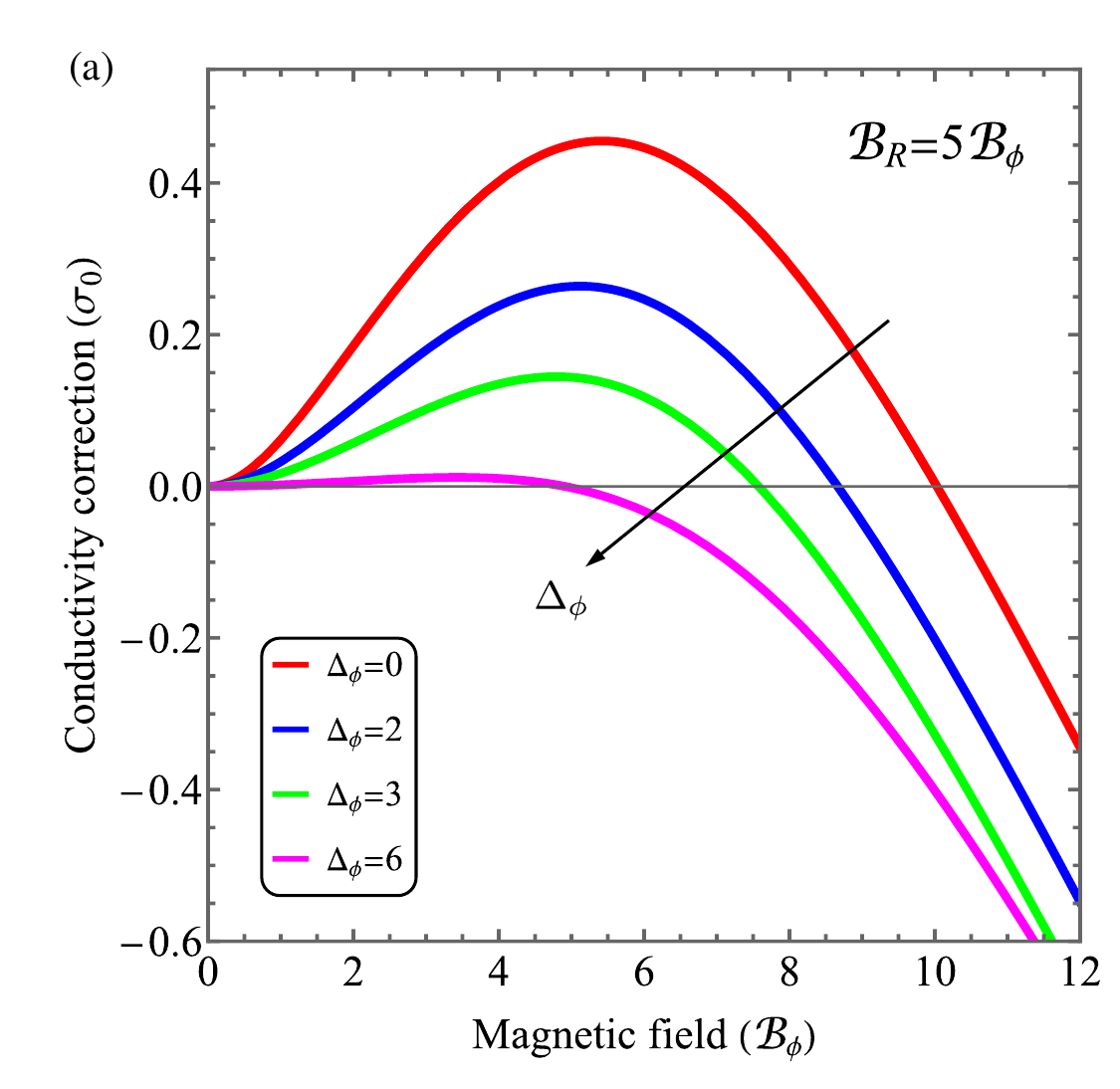} \\
	\centering \includegraphics[width=0.95\linewidth]{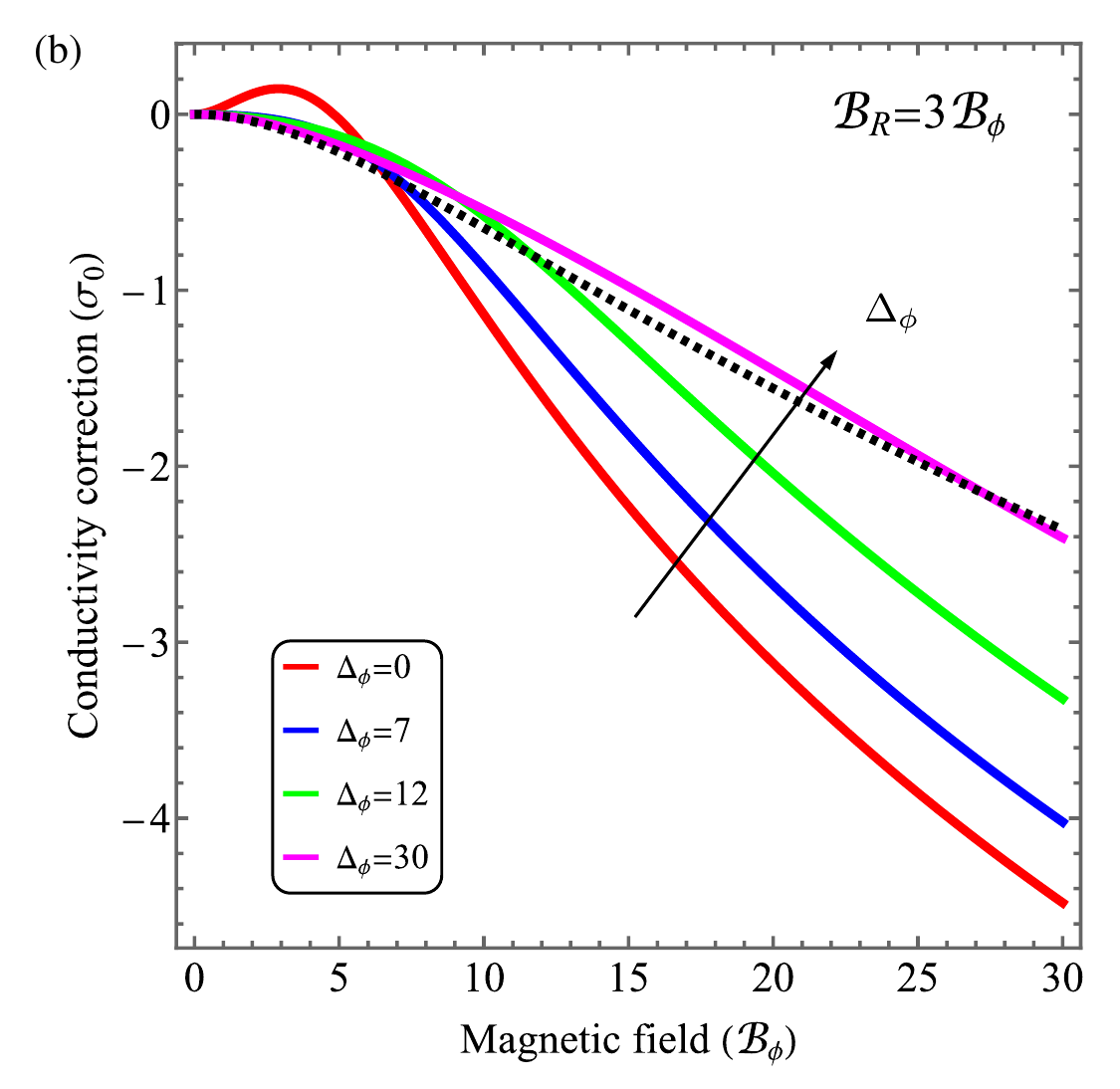} 
	\caption{ 
	Conductivity correction in the absence of intervalley scattering at different values of $\Delta_\phi$ (a) at $\mathcal B_{\rm R}/\mathcal B_\phi=5$ at low fields, (b) at $\mathcal B_{\rm R}/\mathcal B_\phi=3$ in a wider magnetic field range. The dashed curve shows the asymptotic Eq.~\eqref{asympt}.
	}
	\label{Fig_VZ_effect}
\end{figure}

In Fig.~\ref{Fig_VZ_effect} we demonstrate the effect of the valley-Zeeman splitting on WL in Rashba-coupled graphene. At $\lambda_{\rm VZ}=0$, WL is present with $\Delta \sigma(B)>0$ in low fields, see Fig.~\ref{Fig_VZ_effect}(a) and Eq.~\eqref{no_VZ}. The effect of the valley-Zeeman splitting is in the competition with the Rashba splitting. At large  $\lambda_{\rm VZ}$, the eigenstates of the Hamiltonian~\eqref{H} have spins aligned along almost $\pm z$ directions. The system consists of two independent spin subsystems without any effect of spin degrees of freedom. As a result, WAL takes place at 
$\Delta_\phi \geq 6$
as in ordinary graphene without any spin splitting, see Fig.~\ref{Fig_VZ_effect}.

At $\Delta_\phi \gg 1$, the contributions of the $t_0$ and $s$ spin channels coupled by the valley-Zeeman splitting, see Eq.~\eqref{L_matrix}, cancel each other as in the absence of the Rashba splitting. Therefore, the conductivity correction $\Delta\sigma_{\rm intra}$ in each valley is due to spin $t_{\pm 1}$ channels only. They give equal contributions with the Rashba splitting acting as a dephasing only, see Eq.~\eqref{L_t_matrix_epsilon}. Therefore we obtain WAL induced negative magnetoconductivity
\begin{equation}
\label{asympt}
\Delta\sigma \bigr|_{\Delta_\phi \to \infty} = 
-4 \sigma_0 F(b_\phi + b_{\rm R}).
\end{equation}
It follows from Fig.~\ref{Fig_VZ_effect}(b) that this depependence  is achieved at $\Delta_\phi \geq 30$.
 

With allowance for spin- and valley-dependent disorders as well as for Kane-Mele coupling, staggered sublattice potential and trigonal warping, the dephasing rates in the spin channels are different from $\Gamma_\phi$. The formula~\eqref{sigma_no_iv} is generalized in this case by the substitution of the matrix $\mathcal L$ given by Eq.~\eqref{L_matrix} with $\mathcal L + {\rm diag}(b^{t_1}_{t_1},b^{t_1}_{t_0},b^{t_1}_{t_1},b^{t_1}_{s})$. Here $b^{t_1}_{j}$ are related by Eq.~\eqref{b_Gamma} with the dephasing rates $\Gamma^{t_1}_{j}$ listed in the Table in Ref.~\cite{Golub2024}.
In particular, in the presence of valley-dependent (still intra-valley) but spin-independent disorder we have to substitute $\Gamma_\phi \to \Gamma_\phi+\Gamma_z$, and in the opposite case of spin-dependent and valley-independent disorder we have instead of Eq.~\eqref{L_matrix} $\mathcal L + {\rm diag}(b_{\rm SO},b_{\rm asy},b_{\rm SO},0)$. The corresponding rates $\Gamma_i$ and values $b_i$ 
($i=z,{\rm SO},{\rm asy}$)
are given in Refs.~\cite{McCann2012,Golub2024}.

To summarize this section, the effect of valley-Zeeman splitting in the Rashba-coupled graphene is the WL to WAL transition
demonstrated in Fig.~\ref{Fig_VZ_effect}.

\mysection{Effect of inter-valley scattering} 
With allowance for inter-valley scattering, there are 16 interference channels $(vl,sj)$ including both valley ($v$) and spin ($s$) 
triplet and singlet
ones: $l,j=t_1, t_0, t_{-1},s$,
where
$t_0$ and $t_{\pm 1}$ correspond to spin/pseudospin $z$-projection equal to zero or $\pm 1$.
Intervalley scattering 
leads to the additional dephasing in the $(vt_{\pm 1,0},sj)$ 
channels. 
The corresponding rates $\Gamma_{iv}$ and $\Gamma_{*}=\Gamma_z+\Gamma_{iv}$ were introduced in Ref.~\cite{McCann2012}.

Analysis shows that 8 channels $(vt_{\pm 1},sj)$ with any $j$ contribute independently of all other channels. The only difference from the intra-valley scattering case considered above
is the additional dephasing $\Gamma_{*}$.
The $(vt_{0},ss)$ and $(vs,ss)$ channels
both are decoupled from the others and do not depend on the Rashba splitting, see Supplemental Material~\cite{SM}. They do not cancel each other any more due to the intervalley scattering rate $\Gamma_{iv}$.
The rest 6 channels of interference, spin triplets $(vt_{0},sj)$ and $(vs,sj)$ with $j=t_{\pm 1,0}$, are coupled. The corresponding Cooperon equals to $\mathcal L_{iv}^{-1}$, where 
$\mathcal L_{iv}$ is the 6-rank matrix given by
\begin{equation}
\label{M_matrix_SM}
 \mathcal L_{iv}(\epsilon) = \begin{pmatrix}
\mathcal L_{t}+ 2b_{iv}  \mathcal I_3 & i b_{\rm VZ}S_z
\\ i b_{\rm VZ}S_z &  \mathcal L_{t}+\mathcal L_{\rm SO}
\end{pmatrix}.
\end{equation}
Here the first
and the last  three basis states are taken as $(vt_{0},sj)$ and $(vs,sj)$, respectively, with  $j=t_1,t_0,t_{-1}$,
$S_z={\rm diag}(1,0,-1)$, $\mathcal I_3$ is the rank-3 identity matrix, $\mathcal L_{\rm SO}={\rm diag}(b_{\rm SO},b_{\rm asy},b_{\rm SO})$, 
the matrix $\mathcal L_t$ is given by Eq.~\eqref{L_t_matrix_epsilon},
and $b_{iv}$, $b_{\rm SO}$, $b_{\rm asy}$ are related with $\Gamma_{iv}$, $\Gamma_{\rm SO}$, $2\Gamma_{\rm asy}$ by Eq.~\eqref{b_Gamma}.
Hereafter we assume
that the spin dephasing rates are much lower than the intervalley ones 
$\Gamma_{\rm SO, asy} \ll \Gamma_{iv, *}.$ 
This means that the spin dephasing is important in the valley-singlet channels only, where the inter-valley scattering does not result in dephasing.

Inverting the matrix $\mathcal L_{iv}$ 
and calculating the conductivity correction, we obtain the 
magnetoconductivity 
in the following form
\begin{multline}
\label{sigma_iv}
{\Delta\sigma \over \sigma_0} 
= 2{\Delta\sigma_{\rm intra}(b_\phi+b_{*}) \over \sigma_0} 
+ F\qty({b_\phi+2b_{iv} })-F\qty({b_\phi})
\\-   \sum_{m=1}^6 \qty[\tilde{u}_m \psi\qty(1/2 + b_\phi  - \tilde{v}_m)  - \tilde{u}_m^{(0)}\ln{\qty(b_\phi  - \tilde{v}_m^{(0)})}]
\\ + {2{b}_{iv} \over  (b_\phi+b_{\rm R}+{b}_{iv}-1/2)^2 - {b}_{iv}^2+b_{\rm VZ}^2} 
\\ - {2{b}_{iv} \over (b_\phi+b_{\rm R}+{b}_{iv}+1/2)^2 - {b}_{iv}^2},
\end{multline}
where 
the first term, the two rest terms in the first line, and the other lines are the contributions of the above-mentioned 8, 2 and 6 interference channels, respectively.
Here $F(\beta)$ is the HLN function defined after Eq.~\eqref{no_VZ}, $\tilde{v}_{1\ldots 6}$ are the roots of  $\mathcal D_{iv}(\epsilon)$, and
$\tilde{u}_m = {\mathcal N_{iv}(\tilde{v}_m)/\prod_{m'\neq m}(\tilde{v}_m-\tilde{v}_{m'})}$
with $\mathcal N_{iv}(\epsilon)$ and $\mathcal D_{iv}(\epsilon)$ being the polynomials of the 4th and 6th powers, respectively.
They are obtained from the equality
\begin{equation}
{\mathcal N_{iv}(\epsilon)\over \mathcal D_{iv}(\epsilon)} ={\rm Tr}
\qty(\mathcal E_6 \mathcal L_{iv}^{-1}),
\end{equation}
where $\mathcal E_6={\rm diag}(1,1,1,-1,-1,-1)$.

The obtained Eq.~\eqref{sigma_iv} gives the WL-induced magnetoconductivity at an arbitrary relation between the Rashba and valley-Zeeman spin spittings as well as the inter-valley scattering rates. The rates $\Gamma_{\rm R}$, $\Gamma_{\rm VZ}$, $\Gamma_{iv}$, $\Gamma_*$, $\Gamma_{\rm SO}$, $\Gamma_{\rm asy}$ and $\Gamma_\phi$ are independent parameters of the theory. In particular,  Eq.~\eqref{sigma_iv} describes the quantum correction to the conductivity for any values of the parameter $\lambda_{\rm VZ}/(\hbar \Gamma_{iv})$, also beyond the motional-narrowing regime where it is small~\cite{Cummings2017}, 
and an approximate HLN-like formula with the spin dephasing rate was used~\cite{Zihlmann2018}.
%
%
In the absence of the valley-Zeeman splitting we have from Eq.~\eqref{sigma_iv}
\begin{multline}
\label{no_bVZ_iv}
\Delta\sigma\bigr|_{\lambda_{\rm VZ}=0}= 2\Delta\sigma_{\rm intra}(b_\phi+b_{*}) +\Delta\sigma_{\rm intra}(b_\phi+2b_{iv})
\\ -\Delta\sigma_{\rm intra}(b_\phi),
\end{multline}
where $\Delta\sigma_{\rm intra}(b_\phi)$ is given by Eq.~\eqref{no_VZ}.



\begin{figure}[t]
	\centering \includegraphics[width=0.95\linewidth]{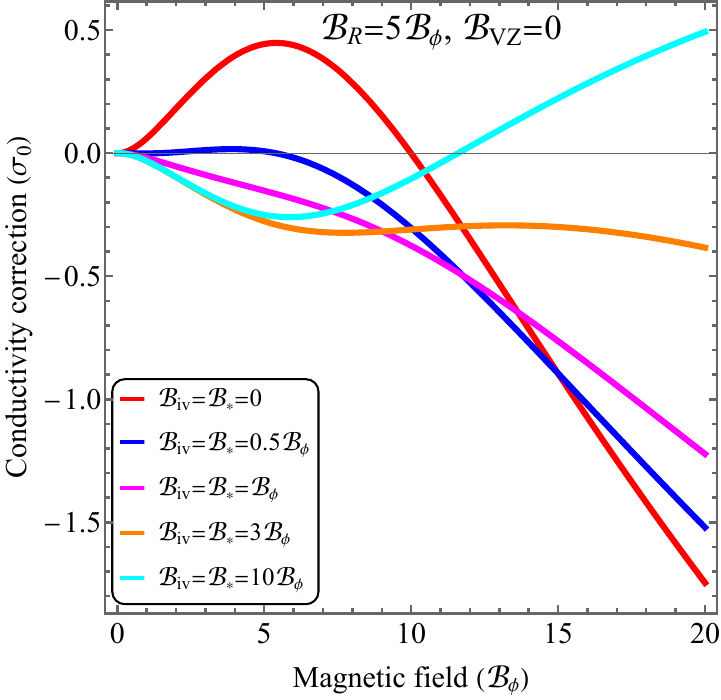} \\
	\includegraphics[width=0.95\linewidth]{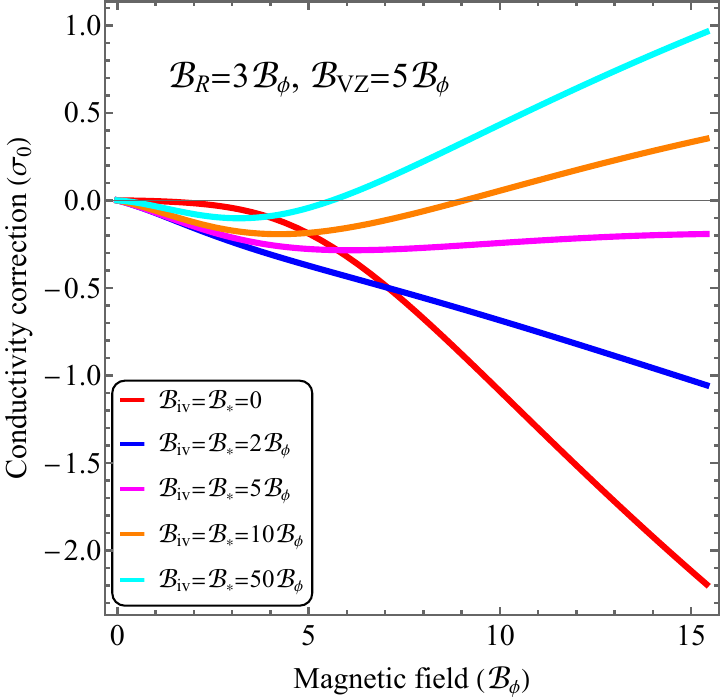} 
	\caption{ 
	Conductivity correction in Rashba-splitted graphene at different intervalley scattering rates. 
	Upper and lower 
	panels correspond to an absence of the valley-Zeeman splitting and to $\mathcal B_{\rm VZ}>\mathcal B_{\rm R}$, respectively.
	The spin-dependent dephasing rates are $\Gamma_{\rm SO}=\Gamma_{\rm asy}=0$.
	}
	\label{Fig_iv-effect}
\end{figure}

Effect of intervalley scattering is demonstrated in Fig.~\ref{Fig_iv-effect}. If the valley-Zeeman splitting is zero, WL takes place in the Rashba-coupled graphene in the absence of inter-valley scattering: the magnetoconductivity is positive in low fields, has a maximum and becomes negative in higher fields. The inter-valley scattering reverses the situation: at large $\Gamma_{iv,*}=10\: \Gamma_\phi$ the magnetoconductivity is negative with a minimum at $B \approx \mathcal B_{\rm R}$, and becomes positive at high fields,  Fig.~\ref{Fig_iv-effect}(a).

A presence of a large valley-Zeeman splitting reverses a situation once more: the magnetoconductivity is negative and its absolute value monotonously increases if the inter-valley scattering is absent, Fig.~\ref{Fig_iv-effect}(b). However, a presence of the inter-valley scattering results in the formation of a minimum of the magnetoconductivity and change of the sign in high fields.

Figure~\ref{Fig_iv-effect} demonstrates  that inter-valley scattering results in the WAL to WL transition in the absence of the valley-Zeeman splitting and to the reciprocal, WL to WAL transition if the valley-Zeeman splitting is large.

\mysection{Conclusion} 
We  showed that both the valley-Zeeman splitting and inter-valley scattering result in an additional phase to the electron interference reversing the sign of the WL-induced magnetoconductivity in the Rashba-coupled graphene.
While the inter-valley scattering results in the sign reversal in any case, the valley-Zeeman splitting plays a role in the presence of the Rashba splitting only. In this case the valley-Zeeman splitting suppresses the effect of the Rashba splitting resulting in the WAL-to-WL or WL-to-WAL transition in the presence/absence of the inter-valley scattering, respectively.

We  derived an analytical expression for the general case of an arbitrary relation between the Rashba and valley-Zeeman splittings as well as inter-valley scattering rates. The limiting cases are  considered where simpler formulas are obtained. The developed theory allows one to determine  adequately the spin- and valley-dependent parameters of graphene heterostructures from experimental data.



\mysection{Acknowledgments}
This work was supported by the Deutsche Forschungsgemeinschaft (DFG, German Research Foundation) via Project-ID 448955585 (Ga501/18).

\bibliography{WAL_Gr_SOI}


\newpage



\onecolumngrid

\newpage
%

\begin{center}
\makeatletter
{\large\bf{Supplemental Material for\\``\@title''}}
\makeatother
\end{center}

\let\oldsec\section

\renewcommand{\thesection}{S\arabic{section}}
\renewcommand{\section}[1]{\oldsec{#1}}
\renewcommand{\thepage}{S\arabic{page}}
\renewcommand{\theequation}{S\arabic{equation}}
\renewcommand{\thefigure}{S\arabic{figure}}

\setcounter{page}{1}
\setcounter{section}{0}
\setcounter{equation}{0}
\setcounter{figure}{0}

\section{Magnetoconductivity  in the absence of intervalley scattering}

In magnetic field, it is convenient to search the Cooperon in the basis of Landau levels of a charge $2e$. Then the conductivity correction reads~\cite{Golub2024}
\begin{equation}
\label{SM_cond_1}
\sigma=2\sigma_0 \qty{ \sum_{n=0}^{N_0} {\rm Tr}\qty[\mathcal E_4 \mathcal C(n)]
+ \mathcal C_0}.
\end{equation}
Here the factor of 2 accounts for two valleys, $n$ enumerates the Landau levels (for the triplet channel the Landau-level numbers are equal to $n+1$), $\mathcal C_0$ is a triplet contribution of the lowest Landau level, $N_0=\mathcal B_{\rm tr}/B \gg 1$ is the cutoff, and the matrix $\mathcal E_4={\rm diag}(1,1,1,-1)$ in the basis of spin triplet and singlet states.

The Cooperon
is given by elements of inverse matrices
\begin{equation}
\mathcal C(n \geq 1) = \qty[\mathcal L(n)]^{-1}, \qquad \mathcal C(n=0) = \mathcal L_0^{-1}, 
\qquad \mathcal C_0=1/( \epsilon_0+b_{\rm R}),
\end{equation}
where
\begin{equation}
\label{L_matrix_n}
\mathcal L(n)=
\begin{pmatrix}
\epsilon_{n-1}+b_{\rm R} & i\sqrt{2 b_{\rm R} n} &0 & 0
\\ -i\sqrt{2 b_{\rm R} n} & \epsilon_n+2b_{\rm R} &i\sqrt{2 b_{\rm R} (n+1)} & -i b_{\rm VZ}
\\ 0 & -i\sqrt{2 b_{\rm R} (n+1)} & \epsilon_{n+1}+b_{\rm R}& 0
\\ 0 & -i b_{\rm VZ} & 0 & \epsilon_n
\end{pmatrix},
\quad
\mathcal L_0=
\begin{pmatrix}
\epsilon_0+2b_{\rm R} &i\sqrt{2 b_{\rm R}}& -ib_{\rm VZ}
\\-i\sqrt{2 b_{\rm R}} & \epsilon_{1}+b_{\rm R}& 0
\\-ib_{\rm VZ}& 0 & \epsilon_0
\end{pmatrix}
\end{equation}
with $\epsilon_n = n+1/2 + b_\phi$.

Calculating 
${\rm Tr}\qty[\mathcal E_4 \mathcal C(n)]$ we obtain
\begin{equation}
\label{sigma_intra_sum_n}
{\sigma(B) \over \sigma_0} = 2 \qty[\sum_{n=0}^{N_0} {\mathcal N(\epsilon_n)\over \mathcal D(\epsilon_n)} - {1\over \epsilon_0+b_{\rm R} -1} + {1\over \epsilon_0+b_{\rm R}}],
\end{equation}
%
where we 
added and subtracted the term ${\rm Tr} \qty{\mathcal E_4 \qty[ \mathcal L(0)]^{-1}}$ to Eq.~\eqref{SM_cond_1} and 
used the relation ${\rm Tr} \qty{\mathcal E_4\qty[ \mathcal L(0)]^{-1}}-{\rm Tr}\qty(\mathcal E_4 \mathcal L_0^{-1})={1/(\epsilon_0+b_{\rm R} -1)}$. 
Here ${\mathcal N(\epsilon_n)\over \mathcal D(\epsilon_n)}={\rm Tr}\qty[\mathcal E_4 \qty[ \mathcal L(n)]^{-1}]$, or explicitly
\begin{equation}
\mathcal N(\epsilon)= 2\qty[\epsilon^3 + 2b_{\rm R}\epsilon^2 + (2b_{\rm R}^2+b_{\rm VZ}^2)\epsilon + b_{\rm R} (b_{\rm R}^2+2b_{\rm R}b_\phi-b_{\rm VZ}^2)],
\end{equation}
\begin{equation}
\mathcal D(\epsilon)=\epsilon^4 + (b_{\rm R}^2+4b_{\rm R}b_\phi-1+b_{\rm VZ}^2) \epsilon^2 +2b_{\rm R} (b_{\rm R}^2+2b_{\rm R}b_\phi+b_{\rm VZ}^2)\epsilon +b_{\rm VZ}^2(b_{\rm R}^2-1).
\end{equation}

The sum~\eqref{sigma_intra_sum_n} can be evaluated owing to the expansion 
\begin{equation}
 {\mathcal N(\epsilon)\over \mathcal D(\epsilon)}
= \sum_{m=1}^4{u_m \over \epsilon - v_m},
\end{equation}
where $v_{1\ldots 4}$ are the roots of  $\mathcal D(\epsilon)$, and
the coefficients $u_{m}$ 
read
\begin{equation}
u_m = {\mathcal N(v_m)\over\prod_{m'\neq m}(v_m-v_{m'})}.
\end{equation}
Calculating the magnetoconductivity $\Delta\sigma=\sigma(B)-\sigma(0)$ we obtain Eq.~(7) of the main text:
\begin{equation}
\label{mcond_intra}
{\Delta\sigma \over \sigma_0} \equiv 2{\Delta\sigma_{\rm intra}(b_\phi) \over \sigma_0} 
= -2 \qty{\sum_{m=1}^4 \qty[u_m \psi\qty(1/2 + b_\phi  - v_m)  - u_m^{(0)}\ln{\qty(b_\phi  - v_m^{(0)})}]
+ {1\over (b_\phi+b_{\rm R} )^2-1/4}},
\end{equation}
where 
$\psi(y)$ is the digamma function.
The coefficients $v_m^{(0)}$ and $u_m^{(0)}$ are the zero-field asymptotes of $v_m$ and $u_m$ calculated by 
passing to the limit $b_{\rm R}^2 \gg 1$ in the function $\mathcal D(\epsilon)$.
%


\section{Magnetoconductivity  in the presence of intervalley scattering}

Intervalley scattering with the rates $\Gamma_{*}$ and $\Gamma_{iv}$ leads to the additional dephasing in the valley channels $vt_1$ and $vt_0$. Accordingly, we introduce four 4-rank matrices
\begin{equation}
\mathcal L_{\pm 1}=\mathcal L(\pm b_{\rm VZ},\Gamma_\phi+\Gamma_{*}),
\quad
\mathcal L_{0}=\mathcal L(0,\Gamma_\phi+2\Gamma_{iv}),
\quad \mathcal L_{s}=\mathcal L(0,\Gamma_\phi),
\end{equation}
where $\mathcal L(b_{\rm VZ},\Gamma_\phi)$ is defined by Eq.~\eqref{L_matrix}.
The matrix of the operator $\mathcal L$ in the basis of 16 states $(lj)$, where $l,j=t_1,t_0,t_{-1},s$ reads
\begin{equation}
\mathcal L=
\begin{pmatrix}
\mathcal L_{1}\\
&\mathcal L_{0} & &\mathcal L_{\rm VZ}
\\ & &  \mathcal L_{-1}
\\ & \mathcal L_{\rm VZ} & & \mathcal L_{s}
\end{pmatrix},
\end{equation}
where $\mathcal L_{\rm VZ}=-i b_{\rm VZ}{\rm diag}(1,0,-1,0)$.

In the above expressions we ignore spin-dependent disorder as well as the Kane-Mele intrinsic spin-orbit coupling, staggered sublattice potential and trigonal warping. If they are present, they result in i)~modifications of the $vt_1$ and $vt_0$ decay rates and ii)~appearance of the rates $\Gamma_{\rm SO}$ and $2\Gamma_{\rm asy}$ in the $(vs,st_1)$ and $(vs,st_0)$ channels, respectively~\cite{Golub2024}. 
With account for the spin-dependent scattering, the matrix $\mathcal L_{s}$ is substituted by  $\tilde{\mathcal L_{s}} = \mathcal L_{s}+{\rm diag}(b_{\rm SO},b_{\rm asy},b_{\rm SO},0)$.
We assume in the following 
that the spin dephasing rates are much lower than the intervalley ones 
$\Gamma_{\rm SO, asy} \ll \Gamma_{iv, *}.$
This means that the spin dephasing is important in the valley-singlet channels only, where the inter-valley scattering does not result in dephasing, and the dephasing rates in the $vt_1$ and $vt_0$ various spin channels have no 
spin-orbit corrections respectively to $\Gamma_{*}$ and $2\Gamma_{iv}$.

The conductivity correction is given by $\sigma= \sigma_0 {\rm Tr}\qty[\mathcal E_s \otimes \mathcal E_v \mathcal L^{-1}]$,
where $\mathcal E_{s,v}={\rm diag}(1,1,1,-1)$ in the basis of the valley or spin triplet and singlet states, respectively.
This yields
\begin{equation}
\sigma= \sigma_0  \sum_{n=0}^{N_0}{\rm Tr}\qty{\mathcal E_s \qty[2\mathcal L_{1}^{-1} 
+ \sigma_z \begin{pmatrix}
\mathcal L_{0} & \mathcal L_{\rm VZ}
\\ \mathcal L_{\rm VZ} &  \tilde{\mathcal L_{s}}
\end{pmatrix}^{-1}]},
\end{equation}
where the factor 2 appears because ${\rm Tr}\qty[\mathcal E_s \mathcal L_{1}^{-1}]={\rm Tr}\qty[\mathcal E_s \qty(\mathcal L_{-1})^{-1}]$.

The matrices $\mathcal L_{0,s,{\rm VZ}}$ have decoupled triplet and singlet sectors:
\begin{equation}
\mathcal L_{0} = \begin{pmatrix}
\mathcal L_{t}+ 2b_{iv}  \mathcal I_3 & 
\\  &  \epsilon_n + 2b_{iv}
\end{pmatrix},
\quad \mathcal L_{s} = \begin{pmatrix}
\mathcal L_{t} & 
\\  &  \epsilon_n
\end{pmatrix},
\quad \mathcal L_{\rm VZ} = -i b_{\rm VZ} \begin{pmatrix}
S_z & 
\\  &  0
\end{pmatrix},
\end{equation}
where $\mathcal L_{t}$ is the triplet part of $\mathcal L_{1}$ at $b_{\rm VZ}=0$ given by Eq.~(5) of the main text, $\mathcal I_3$ is the unit matrix of rank 3, and $S_z={\rm diag}(1,0,-1)$.
Therefore we have:
\begin{equation}
\label{M_matrix}
\sigma= \sigma_0 \sum_{n=0}^{N_0}\qty[ {\rm Tr} \qty(2\mathcal E_4\mathcal L_{1}^{-1} 
+ \mathcal E_6 \mathcal L_{iv}^{-1})
-{1\over \epsilon_n+2b_{iv}} + {1\over \epsilon_n}
], \qquad \mathcal L_{iv} = \begin{pmatrix}
\mathcal L_{t}+ 2b_{iv}  \mathcal I_3 & -i b_{\rm VZ}S_z
\\ -i b_{\rm VZ}S_z &  \mathcal L_{t}+\mathcal L_{\rm SO}
\end{pmatrix},
\end{equation}
where $\mathcal E_6={\rm diag}(1,1,1,-1,-1,-1)$ and $\mathcal L_{\rm SO}={\rm diag}(b_{\rm SO},b_{\rm asy},b_{\rm SO})$.
The last two terms give a contribution to the magnetoconductivity $\Delta\sigma(B)=\sigma_0[F(b_\phi+2b_{iv})-F(b_\phi)]$, where $F(b)=\psi(1/2+b)-\ln{b}$.
The first term equals to $2\Delta\sigma_{\rm intra}(b_\phi+b_{*})$, where the contribution of two individual valleys at pure intra-valley scattering, $2\Delta\sigma_{\rm intra}(b_\phi)$, is given by Eq.~\eqref{mcond_intra}.
As a result, the conductivity correction reads
\begin{equation}
\Delta\sigma= 2\Delta\sigma_{\rm intra}(b_\phi+b_{*}) + \sigma_0   \qty{F(b_\phi+2b_{iv})-F(b_\phi)
+\qty[\sum_{n=0}^{N_0}{\rm Tr}\qty(\mathcal E_6 \mathcal L_{iv}^{-1})  - (B\to 0)]
}.
\end{equation}

In the absence of intervalley scattering when $b_{iv}=b_{*}=b_{\rm SO}=b_{\rm asy}=0$, the matrix $\mathcal L_{iv}$ gives no contribution (because it does not contain $\mathcal E_6$, an analog of $\sigma_z$), and we get $\Delta\sigma(b_{iv,*,{\rm SO},{\rm asy}}=0)= 2\Delta\sigma_{\rm intra}(b_\phi)$.

In the absence of the valley-Zeeman splitting when $b_{\rm VZ}=0$,  we have 
$\Delta\sigma_{\rm intra}=\sigma_0[F - \mathcal F_t]$.
Therefore we obtain
\begin{equation}
\label{zero_bVZ}
\Delta\sigma\bigr|_{b_{\rm VZ}=0}= 2\Delta\sigma_{\rm intra}(b_\phi+b_{*}) +\Delta\sigma_{\rm intra}(b_\phi+2b_{iv})-\Delta\sigma_{\rm intra}(b_\phi,b_{\rm SO},b_{\rm asy}),
\end{equation}
where we 
took into account that $\mathcal L_t^{-1}$ yields the contribution $-\sigma_0 \mathcal F_t(b_\phi,b_{\rm R})$ and 
$(\mathcal L_{t}+\mathcal L_{\rm SO})^{-1}$
yields $-\sigma_0 \mathcal F_t(b_\phi,b_{\rm R},b_{\rm SO},b_{\rm asy})$. 
Here the function $\mathcal F_t$ is given by~\cite{Golub2024}
\begin{equation}
\mathcal F_t(b_{\rm R},b_\phi) = \sum_{m=1}^3 \qty[y_m\psi(1/2+b_\phi-w_m)-y_m^{(0)}\ln(b_\phi-w_m^{(0)})],
\quad
{\rm Tr}\qty[\mathcal L_t^{-1}(\epsilon)] = \sum_{m=1}^3 {y_m \over \epsilon - w_m},
\end{equation}
and $y_m^{(0)}$, $w_m^{(0)}$ are the zero-field asymptotes of $y_m$, $w_m$.
Explicit expressions for $y_m$ and $w_m$ are given in Ref.~\cite{Golub2024}.

In the general case, we get
\begin{multline}
\sum_{n=0}^{N_0}{\rm Tr}\qty(\mathcal E_6 \mathcal L_{iv}^{-1})= 
\sum_{n=0}^{N_0} 
{\mathcal N_{iv}(\epsilon_n)\over \mathcal D_{iv}(\epsilon_n)}
- {\rm Tr}\qty[\sigma_z \begin{pmatrix}
b_\phi+b_{\rm R}-1/2+ 2b_{iv} & -i b_{\rm VZ}
\\ -i b_{\rm VZ}&  b_\phi+b_{\rm R}-1/2
\end{pmatrix}^{-1}]
+{1\over \epsilon_0+b_{\rm R} +2b_{iv}}- {1\over \epsilon_0+b_{\rm R}}
\\= 
\sum_{n=0}^{N_0} 
{\mathcal N_{\rm VZ}(\epsilon_n)\over \mathcal D_{\rm VZ}(\epsilon_n)}
+2b_{iv}\qty[{1\over (b_\phi+b_{\rm R}+b_{iv}-1/2)^2 - b_{iv}^2+b_{\rm VZ}^2}-{1\over (b_\phi+b_{\rm R}+b_{iv}+1/2)^2 - b_{iv}^2}],
\end{multline}
where $\mathcal N_{iv}(\epsilon)$ and $\mathcal D_{iv}(\epsilon)$ are the polynomials of the 4th and 6th powers, respectively.
They are obtained from the equality
\begin{equation}
{\mathcal N_{iv}(\epsilon)\over \mathcal D_{iv}(\epsilon)} ={\rm Tr}\qty[\mathcal E_6 \mathcal L_{iv}^{-1}(\epsilon)],
\end{equation}
where $\mathcal L_{iv}(\epsilon)$ is given by Eq.~\eqref{M_matrix} with $\mathcal L_t(\epsilon)$ equals to the triplet sector of the matrix $\mathcal L(n)$ from Eq.~\eqref{L_matrix_n} with $n=\epsilon -b_\phi-1/2$ (and $\epsilon_{n}=\epsilon$, $\epsilon_{n\pm 1}=\epsilon\pm 1$):
\begin{equation}
\mathcal L_{iv}(\epsilon) = \begin{pmatrix}
\epsilon-1+b_{\rm R}+ 2b_{iv} & i\sqrt{2 b_{\rm R} n} &0 &  -i b_{\rm VZ}&0 &0 
\\ -i\sqrt{2 b_{\rm R} n} & \epsilon+2b_{\rm R}+ 2b_{iv} &i\sqrt{2 b_{\rm R} (n+1)} &0&0 &0 
\\ 0 & -i\sqrt{2 b_{\rm R} (n+1)} & \epsilon+1+b_{\rm R}+ 2b_{iv}&0 &0 &i b_{\rm VZ}
\\ -i b_{\rm VZ}&0 &0 &  \epsilon-1+b_{\rm R}+b_{\rm SO}& i\sqrt{2 b_{\rm R} n}&0
\\ 0&0 &0 &  -i\sqrt{2 b_{\rm R} n} & \epsilon+2b_{\rm R}+b_{\rm asy} &i\sqrt{2 b_{\rm R} (n+1)}
\\ 0 &0 &i b_{\rm VZ} &0 & -i\sqrt{2 b_{\rm R} (n+1)} & \epsilon+1+b_{\rm R}+b_{\rm SO}
\end{pmatrix}.
\end{equation}

Then we use the expansion
\begin{equation}
{\mathcal N_{iv}(\epsilon)\over \mathcal D_{iv}(\epsilon)} = \sum_{m=1}^6{\tilde{u}_m \over \epsilon - \tilde{v}_m},
\end{equation}
where $\tilde{v}_{1\ldots 6}$ are the roots of  $\mathcal D_{iv}(\epsilon)$, and
the coefficients $\tilde{u}_{m}$ read $\tilde{u}_m = {\mathcal N_{iv}(\tilde{v}_m)/\prod_{m'\neq m}(\tilde{v}_m-\tilde{v}_{m'})}.$
This yields Eq.~(13) of the main text:
\begin{multline}
\Delta\sigma= 2\Delta\sigma_{\rm intra}(b_\phi+b_{*}) + \sigma_0   \qty[F(b_\phi+2b_{iv})-F(b_\phi)]
+ \sigma_0   \Biggl\{ -\sum_{m=1}^6 \qty[\tilde{u}_m \psi\qty(1/2 + b_\phi  - \tilde{v}_m)  - \tilde{u}_m^{(0)}\ln{\qty(b_\phi  - \tilde{v}_m^{(0)})}]
\\
+2b_{iv} \qty[ {1\over  (b_\phi+b_{\rm R}+b_{iv}-1/2)^2 - b_{iv}^2+b_{\rm VZ}^2} - {1\over (b_\phi+b_{\rm R}+b_{iv}+1/2)^2 -b_{iv}^2}]
\Biggr\}.
\end{multline}
%
Again, at $b_{\rm VZ}=0$ we have from the terms in curly brackets $-\sigma_0[\mathcal F_t(b_{\rm R},b_\phi+2b_{iv})-\mathcal F_t(b_\phi,b_{\rm R},b_{\rm SO},b_{\rm asy})]$, and together with 
the other terms
this gives Eq.~\eqref{zero_bVZ}.

Substitution of $2\Delta\sigma_{\rm intra}$ given by Eq.~\eqref{mcond_intra} yields finally
\begin{multline}
{\Delta\sigma \over \sigma_0} 
= -2 \qty{\sum_{m=1}^4 \qty[u_m \psi\qty(1/2 + b_\phi+b_{*}  - v_m)  - u_m^{(0)}\ln{\qty(b_\phi+b_{*}  - v_m^{(0)})}]
+{1\over (b_\phi+b_{*}+b_{\rm R} )^2-1/4}}
\\+ F(b_\phi+2b_{iv})-F(b_\phi)
-   \sum_{m=1}^6 \qty[\tilde{u}_m \psi\qty(1/2 + b_\phi  - \tilde{v}_m)  - \tilde{u}_m^{(0)}\ln{\qty(b_\phi  - \tilde{v}_m^{(0)})}]
\\ +2b_{iv}\qty[ {1\over  (b_\phi+b_{\rm R}+b_{iv}-1/2)^2 - b_{iv}^2+b_{\rm VZ}^2} - {1\over (b_\phi+b_{\rm R}+b_{iv}+1/2)^2 - b_{iv}^2}]
.
\end{multline}

\end{document}